\documentclass[12pt,final,letterpaper]{iopart}
\usepackage{cite}
\usepackage{graphicx}

\begin{document}

\title{All-atomic source of squeezed vacuum with full pulse-shape control}
%Quadrature-noise pulse shaping of a squeezed vacuum?
%Noise pulse shaping in a squeezed vacuum using magnetic field?
%Pulse control over the quadrature noise of squeezed vacuum using magnetic field?
%Pulse-shape quantum noise control in Rb atoms with magnetic field
%Pulse control of squeezed vacuum via polarization self-rotation

\author{Travis Horrom, Irina Novikova, Eugeniy E. Mikhailov$^*$}

\address{College of William \& Mary, Williamsburg, VA. 23185, USA
\\
$^*$Corresponding author: eemikh@wm.edu}
\begin{abstract}

We  report on the generation of  pulses  of a low-frequency squeezed  vacuum
with noise suppression $>2$dB  below the standard quantum limit in a
hot resonant ${}^{87}$Rb  vapor with polarization
self-rotation. We  demonstrate the possibility to precisely control
the temporal  profile of the squeezed noise quadrature by applying a calibrated longitudinal
magnetic field, without degrading the maximum amount of squeezing.

\end{abstract}

%Squeezed states, 42.50.Dv
%Phase coherence quantum optics, 42.50.Gy
%Self-phase modulation (nonlinear optics), 42.65.Jx
\vspace{2pc}
\submitto{\JPB}

\maketitle

%\section{Introduction}

Growing interest in practical realizations of quantum information technology
stimulates a broad exploration of various optimal protocols and
infrastructures. One of the leading approaches is based on using optical
fields as quantum information (QI) carriers that can be strongly coupled
with  resonant matter systems, such as atoms (warm or
cold)~\cite{kimbleNature08, lvovskyNPh09, polzikRMP10, euromemory,
novikova2012review},  ``atom-like''
defects in solid-state systems (such as nitrogen vacancy centers in
diamond~\cite{Dutt01062007}), and nanostructures (such as quantum dots~\cite{PhysRevLett.89.207401}). While many initial QI
protocols relied on qubits formed by individual single photons, continuous
variable approaches have become a promising alternative~\cite{polzik_book,andersenLPR2010}. In
this manuscript we demonstrate an important tool for continuous variable
quantum information, namely a source of pulsed squeezed vacuum  with
reliable and simple control over the temporal output mode.

Any experimental implementation of continuous variable measurements or
operations requires precise knowledge, or better yet, active control over
the spatial and temporal profile of the involved optical fields. It
becomes important, for example, for reliably reconstructing the quantum
states~\cite{lvovskyRMP09}, and for matching the bandwidth of an optical
signal with the linewidth of a resonant light-atom interaction to achieve
optimal coupling~\cite{polzikRMP10}. Good examples of the latter use of temporal pulse
shaping can be found in  realizations of maximally efficient quantum memory in atomic
ensembles~\cite{lvovskyNPh09,novikovaLPR11}. Recent comprehensive theoretical studies
by Gorshkov \emph{et al.} considered a wide range of potential quantum
memory
protocols~\cite{gorshkovPRL,gorshkovPRA1,gorshkovPRA2,gorshkovPRA3,gorshkovPRA4},
such as: electromagnetically induced transparency (EIT) in a cavity and in free
space, far-off-resonant Raman, and a variety of spin echo methods including
ensembles enclosed in a
cavity, inhomogeneous broadening, and high-bandwidth non-adiabatic storage.
Theoretically, high optical depth is necessary to achieve a storage
efficiency close to $100~\%$ for most of these memories~\cite{gorshkovPRL}.
In practice however, residual absorption and competing nonlinear processes
make atomic ensembles with moderate optical depths most practical. In these situations, shaping of quantum optical
signals into predetermined temporal envelopes may be required to achieve
optimal efficiency. For example, for EIT-based quantum memories, the
efficiency of quantum storage for a given optical depth is fundamentally limited by the balance
between the compression of an optical pulse inside the limited length
of an atomic ensemble, and the width of the transparency spectral window.
Thus, the temporal profile of signal and/or control optical
fields must be tailored to minimize losses and store the signal with
optimal efficiency~\cite{novikovaPRL07opt,novikovaPRA08,duOL11}.

%\textbf{think of more applications}

Traditionally, squeezed light is produced via parametric down-conversion in
nonlinear crystals~\cite{bachor_guide_2004}. The development of proper periodic poling
masks for KTiOPO$_4$ crystals (PPKTP) enabled efficient quasi-phase matched
down-conversion around 800~nm, and made possible the development of new generation of
solid-state sources for single photons and continuous squeezed and entangled fields 
at the frequency of the Rb D${}_1$ line.
However, a typical spectral bandwidth of the nonlinear optical conversion in such a crystal
is rather broad (around a few nm).  At the same time, the bandwidths of many light-atom
coherent interactions, such as EIT or Raman resonances, do not often
exceed a few MHz.  For efficient interfacing with these systems, the squeezing
sources need a
significant reduction of technical noise at these low sideband
frequencies. While it is possible to achieve squeezing at sub-MHz frequencies
in nonlinear crystals using a combination of narrow-band powerful pump lasers, high-quality cavities, and
sophisticated electronics for active stabilization and
feedback~\cite{hetet_squeezed_at_D1_Rb_2007,furusawaOE07,giacobinoOE2009}, this remains a very challenging technical task, and requires
a significant amount of resources and lab space. Nevertheless, PPKTP-based sources of squeezed vacuum have been successfully integrated
with quantum memories in Rb atoms: both under EIT conditions in both cold
and hot Rb
atoms~\cite{lvovsky08prl_sq_eit,furusawa08_prl_sq_eit,lvovsky09njp_squeeziong_eit,kozumaPRA10},
and for the far off-resonance Raman interactions~\cite{polzikNatureP11}.
Generation of pulsed squeezed vacuum fields using parametric down conversion-based squeezing methods is also
non-trivial. Since these cavity-based crystal squeezers operate in the CW regime and
generate a continuous squeezed vacuum field, the formation of the pulses has been
done externally, using either mechanical choppers or acousto-optical
modulators (AOM). Either of these approaches has serious drawbacks: an additional
optical element such as an AOM introduces additional losses, and may distort
the spatial mode of the generated field due to thermal nonhomogeneities in
its nonlinear crystal. Mechanical choppers, such as rotating slits, do
not add any losses, but produce only rectangular temporal envelopes of
fixed duration, which are not ideal for many experiments.

As an alternative to crystal squeezers, optical nonlinearities in resonant atomic
media can lead to the generation of
non-classical optical fields as well. For example, two-mode squeezed and
entangled optical fields were produced as a result of non-degenerate
four-wave
mixing~\cite{slusherPRL85,shahriarOC98,lettSci08,lettPRA08,polzik2009oe}
with capabilities of a  nanosecond long squeezed pulse generation~\cite{grangier2011NJP}.
A similar interaction for the
degenerate case leads to the generation of a resonant broad-band
quadrature-squeezed vacuum optical field~\cite{matsko_vacuum_2002}, commonly
referred to as polarization self-rotation (PSR) squeezing. This
method relies on the strong interaction between two linear polarizations of a
near-resonant optical field propagating through an ensemble of resonant atoms.
In the case of an elliptically polarized input, the differential light shifts
of various Zeeman transitions result in circular birefringence of the
atomic vapor and in the rotation of the polarization ellipse by an angle
proportional to the input
ellipticity~\cite{davisOL92,budkerPRA01,novikova_large_sr_squeezing_2002,mikhailov_psr_mot2011}.
An input linear polarization is not rotated, but the same nonlinear interaction is
manifested in strong cross-phase modulation between the two orthogonal
components (the strong input laser field and the orthogonal vacuum field).
Even though the pump laser field propagates through the atoms unaffected,
the orthogonally polarized vacuum field becomes quadrature
squeezed~\cite{ries_experimental_2003,mikhailov2008ol,mikhailov2009jmo,grangier2010oe,lezama2011pra}.
This method is particularly attractive for applications involving
narrow-band coherent light-atom processes, since quantum noise suppression
at noise sideband frequencies down to 20~kHz has been recently
demonstrated~\cite{mikhailov2008ol}. Other advantages of this method
include the relative simplicity of the experimental setup, moderate
requirements for pump laser power (a few mW), automatic matching of the
wavelength of the generated squeezed field to the corresponding atomic
transition wavelength, and the possibility to produce pulses of squeezed
light by modulating the intensity of the pump laser~\cite{grangier2010oe}.

In this paper, we investigate the modification of quantum fluctuations of a
vacuum optical field after propagating through a Rb vapor under PSR
conditions in the presence of a magnetic field. 
In this experiment, we demonstrate that a
longitudinal magnetic field allows for good control over the temporal profile
of the generated squeezed field, and thus such a setup can be used as a source of
pulsed squeezed vacuum with programmable pulse shapes.  This method offers
important advantages since it does not change the angle of the squeezing
quadrature with respect to the local oscillator, and thus allows convenient
pulse shape formation without the need to adjust this angle as well. Also, a
modest magnetic field does not affect the strong pump field. Thus, this
field can be used afterwards as a local oscillator that intrinsically has
the same spatial profile as the squeezing beam.

%To provide some intuitive explanation of the magnetic field effect on the
%quadrature noise, let us briefly consider propagation of two linearly
%polarized optical fields -- a strong $y$-polarized pump field (treated
%classically) $E_y$, and a weak $x$-polarized field $E_x$-- through an
%atomic medium, as it was previously considered in
%~\cite{matsko_vacuum_2002}. In this model the rotation of the polarization
%ellipse is caused by a phase difference acquired by the two circular
%components caused by differential light-shift on different Zeeman sublevels
%$\phi_{sr}(z)=\epsilon(z) \mathrm{g}
%z$~\cite{novikova_large_sr_squeezing_2002}, where $\mathrm{g}$ is a
%phenomenological self-rotation coefficient, and $\epsilon$ is ellipticity
%of the overall optical field $\epsilon \simeq -i (E_x-E^*_x)/E_y$. In the
%presence of a longitudinal magnetic field there is an additional phase
%$\phi_B \propto B$~\cite{Xxx}, which in the limit of small $\epsilon$ does
%not depend on the ellipticity of the light~\cite{novikovaellNMOR}. These
%two independent phase contributions add up together, leading to the
%following evolution for the output fields $E_x(L)$ and $E_x(L)$:
%%
%\begin{eqnarray} \label{phenom5}
%E_x(L) &=& E_x(0)-i g L(E_x(0)-E_x^*(0)) + \frac12 \phi_B E_y(0), \\
%E_y(L) &=& E_y(0).
%\end{eqnarray}
%in the relevant approximation approximation $\phi_{sr},\phi_B \ll 1$.

\section{Experimental arrangement}

\begin{figure}[h]
	\begin{center}
	\includegraphics[width=0.7\textwidth]{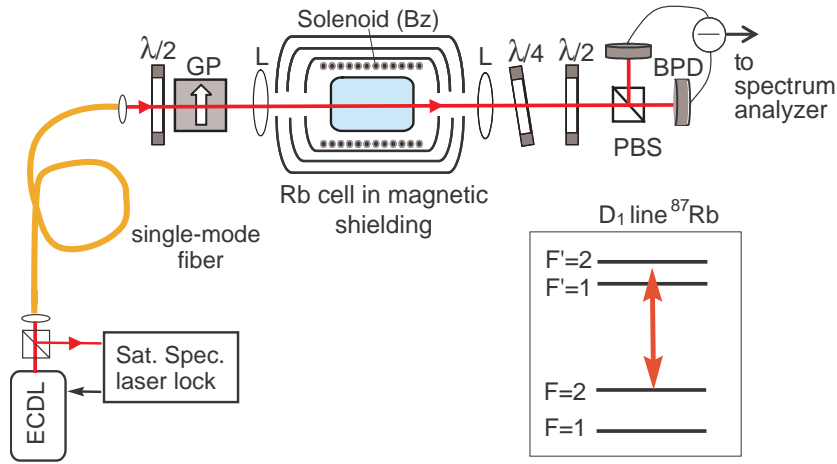}
	\end{center}
	\caption{Experimental setup. The description of components is provided in
	the text.}
	\label{fig:setup}
\end{figure}

The experimental setup  is depicted in Fig.~\ref{fig:setup}.  The output of
an external cavity diode laser (ECDL) is tuned to the $^{87}$Rb D$_1$ line
$F_g=2 \to F_e=2$ transition  and actively locked  to this transition using a saturation
spectroscopy  dither lock.  The laser output is sent through a polarization-maintaining
single mode fiber to clean its spatial mode, achieving an axially symmetric
Gaussian intensity distribution.  Then light passes through a
half-wave  plate  and Glan-Laser polarizer (GP) combination,  which serves  as  a
power attenuator  and, most importantly,  produces a high  quality linearly
polarized pump laser beam.  The laser beam is  focused with a lens ($f=400$ mm)  to achieve a
100~$\mu$m beam waste
approximately in the center of the  75~mm-long Pirex cell with isotopically
pure $^{87}$Rb. The cell is maintained
at 66$^\circ$ Celsius, corresponding to an atomic number density of
$5.4\times10^{11}$ atoms/cm$^3$, 
which we found experimentally to give the best noise suppression in squeezed vacuum.  
The cell is surrounded by a three layer $\mu$-metal
magnetic shield and placed inside a  solenoid which gives  us precise
control  over the internal longitudinal  magnetic  field.  After the  cell,  we
collimate the  laser beam with  a second lens ($f=300$ mm). For squeezing detection,
the  beam polarization is rotated  by 45$^\circ$
with respect  to the axis of a polarizing  beam splitter (PBS) using another
half-wave  plate,  to  achieve  a 50/50  splitting ratio.  After  the  PBS, the two
split  laser beams  are directed  to a  balanced photo  diode (BPD).  Such a
combination  of PBS  and  BPD  makes a  very  stable  homodyne detector,  because
the  phase  between the linearly  polarized  pump  beam (used  as
local oscillator  (LO)) and the orthogonally  polarized squeezed vacuum is
fixed  and  both fields  share  the
same path.  This setup  was described in~\cite{lezama2011pra}  and provides
superior  homodyne  phase  stability  compared to  our  previous  detection
scheme~\cite{mikhailov2008ol, mikhailov2009jmo} which  separated the LO and
squeezing beam paths. Additionally, the overlap of the squeezing beam and LO
is 100\% since they are essentially the same beam and travel together until
the  last  PBS.  This  increases  detection efficiency  and  has allowed
squeezing  levels of up to  3~dB in similar  setups~\cite{lezama2011pra}. The
output of the  BPD goes to a  spectrum analyzer (SA). The fine  control over the
measured noise  quadrature angle is  obtained with a quarter-wave  plate, which
is placed after the collimating lens and set in  such a way that the ordinary and extraordinary axes  coincide with the
laser  beam  polarizations.  In this arrangement, a small  tilt of  the  quarter-wave plate  introduces  a
controllable phase shift between the squeezed vacuum and LO beam.

\begin{figure}[h]
	\begin{center}
	\includegraphics[width=0.6\textwidth]{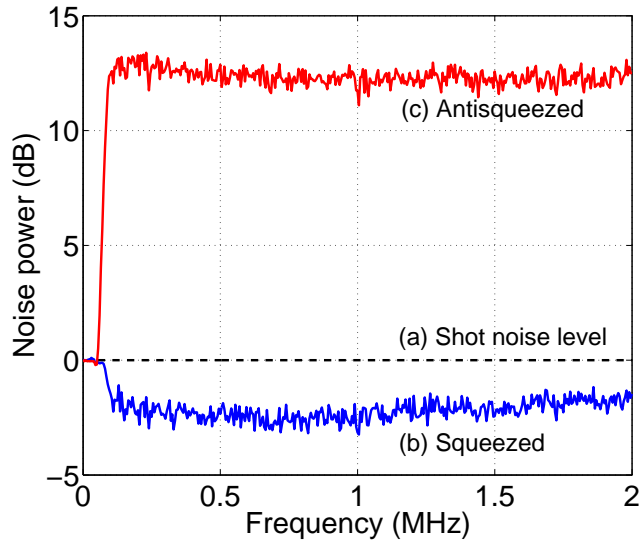}
	\end{center}
	\caption{Typical noise spectrum of squeezed and anti-squeezed quadratures.  Laser pump
	power~$= 7.0$~mW. RBW $= 10$ kHz.}
	\label{fig:noise_spectrum}
\end{figure}

In the absence of magnetic field, we observe a suppression of the quantum noise in the
vacuum field below the standard quantum limit (shot  noise). This
suppression exceeds  2~dB,  and  spans  from  around  100 kHz to
several MHz  (see Fig.~\ref{fig:noise_spectrum}).  We
calibrate to the shot noise  level by  inserting a  polarizer after  the cell,
which  completely rejects the squeezed vacuum  field and  transmits only  the LO
field. Weak absorption in this polarizer  decreases the shot noise level by
about  0.2~dB, consequently  the  actual/corrected amount  of squeezing  is
higher by this number. The total optical loss in the squeezed vacuum path
is equal to 10\% (3\% at the output window of the cell, 7\% at the
steering mirrors, collimating lens, and balancing PBS); we also estimate the quantum
efficiency of our photo diodes to be 95\%. Taking into account all of these
factors we infer that our squeezer produces around 3.6~dB of squeezing,
from which we are able to detect only 2.3~dB due to losses.
Though the inferred 3.6~dB value is
still lower than the theoretically predicted 6~dB by Matsko {\it et
al.}~\cite{MatskoNWBKR02}, our squeezer is reasonably close to predicted
values, and one expects to see even higher squeezing with cold atoms in a
magneto-optical trap~\cite{mikhailov2011jmo}.
We choose to report only the experimentally
measured values of squeezing and do  not use the above corrections (even the
shot noise reduction) anywhere in the manuscript.

One possible way to modulate the noise level of the PSR squeezing
is by controlling the input  pump power. In particular, lowering the pump power degrades and eventually
kills the squeezing process,
and thus makes it possible to shape the noise pulses~\cite{grangier2010oe}. However, this variation of the pump laser
intensity makes it difficult to reuse the same beam as a
LO and accurately measure the quantum noise levels.
This problem  can  be alleviated  by using an independent LO  field, but at the price of the
increased complexity of the experiment, a lower quantum detection efficiency due
to the imperfect overlap of the LO and squeezing beams, and a reduced homodyne
phase stability.  In addition, we note that one cannot modulate the light
signal after it is already squeezed due to the resulting degradation of the noise
suppression.

\begin{figure}[h]
	\begin{center}
	\includegraphics[width=0.5\textwidth]{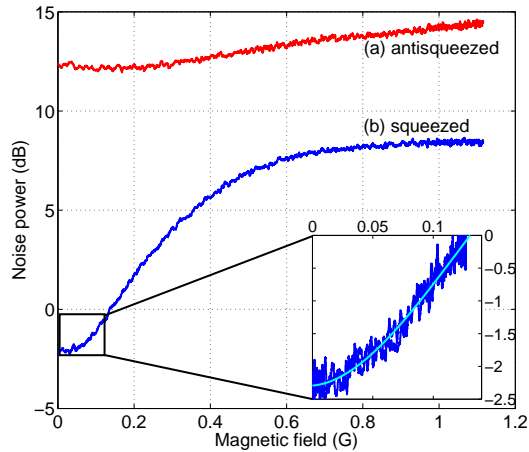}
	\end{center}
	\caption{Noise power (dB) of (a)anti-squeezed and (b)squeezed quadratures vs
	applied longitudinal magnetic field.  Inset zooms in on small fields used
	for this experiment. The spectrum analyzer is set to 1~MHz central
	frequency with the RBW=100~KHz.}
	\label{fig:noise_vs_b}
\end{figure}

Here we  demonstrate a novel approach to squeezing manipulation which avoids
these drawbacks. We  modify the quantum
noise quadrature level resulting from the squeezing process by  changing the  longitudinal magnetic  field
($B_z$)  applied with  the  solenoid. Fig.~\ref{fig:noise_vs_b} shows measurements
of the minimum (squeezed) and maximum (antisqueezed) noise quadratures as functions
of the applied longitudinal magnetic field.  It is clear that the  squeezed
quadrature has a strong
dependence on the  magnetic field while the  anti-squeezed quadrature noise
level has a much weaker response. Similar increase of intensity noise
with magnetic field have been previously reported for weak optical
fields~\cite{sautenkov2008josab}. By
modifying  $B_z$,  we can  change  the  squeezing level  from a
maximally  squeezed state  to  the  shot noise  limited  level (and beyond).
Thus, the  $B_z$
dependence on time translates to a squeezing noise time-dependence. Both the
squeezing  and anti-squeezing noise levels
have a nonlinear response to the longitudinal magnetic field (see
Fig.~\ref{fig:noise_vs_b}) which must be taken into account
for the generation of an arbitrary  pulse shape.
%The control of the magnetic field pulse  shapes is achieved via voltage controlled current
%source,  which  is  driven  by programmable  SRS  DS345  function/arbitrary
%waveform generator.

There are several remarkable features of the observed magnetic field dependence. First, we verified that the change in minimum noise quadrature caused by the applied magnetic field is not accompanied by any changes in its phase with respect to the local oscillator. This means that any squeezing pulses formed by changing the magnetic field will have a uniform phase with no chirping.  We also note that the transverse magnetic field has a much weaker effect on  the squeezed quantum  noise. In particular, we observed no measurable deterioration of quantum noise suppression
due to transverse magnetic fields up to several Gauss, even though this is a much stronger field than we apply in the longitudinal direction.

To gain some intuitive explanation of the magnetic field effect on the
quadrature noise, one has to consider the effects of Zeeman shifts of the relevant magnetic sublevels.
In the simplified model developed in~\cite{matsko_vacuum_2002}, the self-rotation of the elliptical polarization
is caused by a circular birefringence induced by a differential light-shift due to unbalanced intensities of two circular polarizations.
For a linearly polarized pump field, which can be decomposed into two circular components of equal intensity, all the Zeeman sublevels are shifted equally, and the self-rotation mechanism serves as a ``quantum feedback'' for quantum fluctuations. Additional longitudinal magnetic field breaks the degeneracy of the magnetic sublevels, producing non-zero phase shift between two circular components and, as a result, the nonlinear Faraday rotation of the original polarization~\cite{budkerRMP02}. It is important to note that the overall rotation of the linear pump polarization due to this effect is rather small, and does  not modify  the shot  noise level for small ($B<0.2$~G) magnetic fields. We also verified that this rotation cannot explain the deterioration of squeezing since it cannot be compensated for by optimizing the angles of the half and quarter waveplates at the detection stage. At the same time, the difference in acquired phases for different spectral components of the vacuum field results in a non-stationary minimum quadrature angle, and leads to an overall reduction of detectable squeezing.
This simple model may also explain the weak effect of the transverse magnetic field. For this so called Voight configuration, the polarization rotation is caused by linear dichroism of the atomic medium~\cite{budkerRMP02}. This effect is quadratic in Zeeman shift, and thus is significantly weaker than the rotation in the Faraday configuration.

\section{Quadrature-noise pulse shaping}
\begin{figure}[h]
	\begin{center}
	\includegraphics[width=1.0\textwidth]{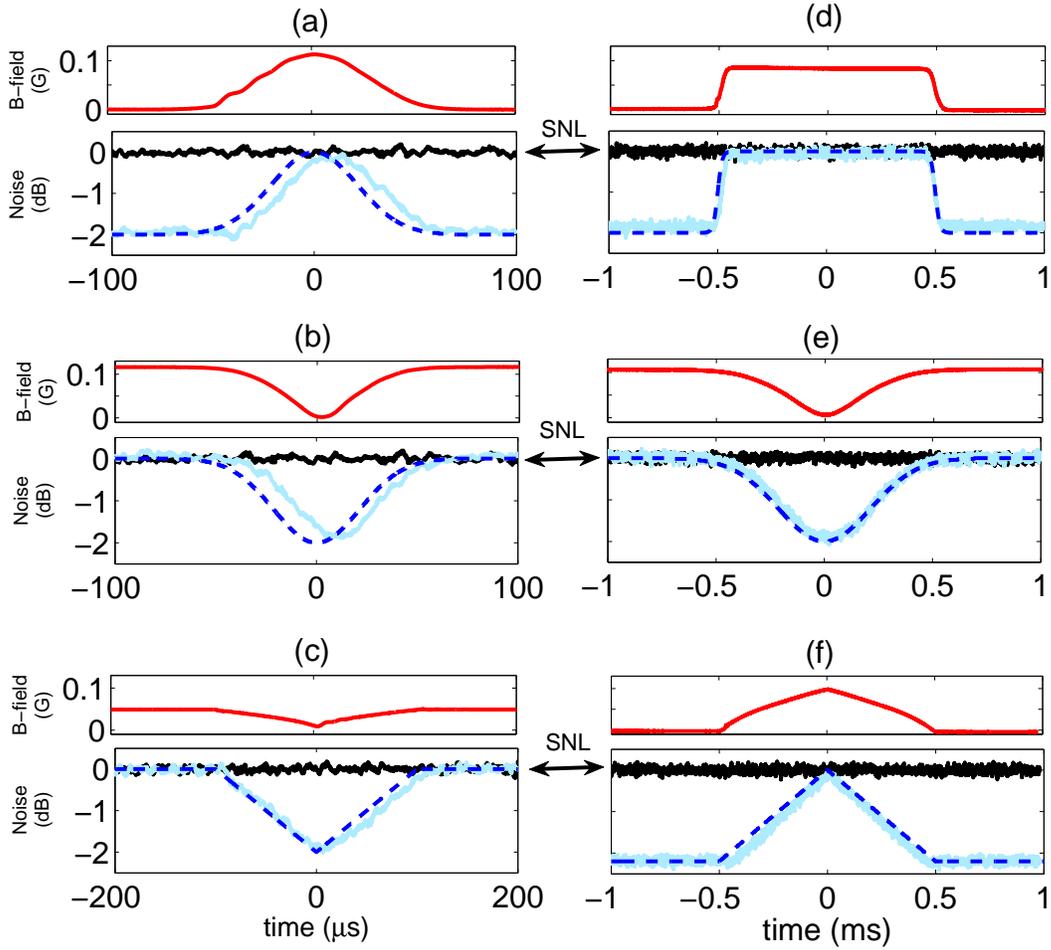}
	\end{center}
	\caption{
		Modulation of the quantum noise with different pulse shapes.  Top plots:
		magnetic field pulses applied to the atoms.  Bottom plots: resultant
		squeezed noise pulses compared with the shot noise limit (SNL).
		Desired noise pulse shapes shown with dashed lines were: Gaussian (a, b, and e), triangular (c and f), and square (d).
		\label{fig:pulses}
	}
\end{figure}

To    demonstrate  the capabilities    of    our   method,    we chose  several different temporal profiles with different parameters for the output squeezed vacuum field. Fig.~\ref{fig:pulses} shows a few example pulse shapes: (a) a positive Gaussian pulse of around $30 \mu s$ duration, (b) a $60 \mu s$ negative Gaussian pulse, (c) a negative $200 \mu s$ triangular pulse, (d) a positive 1 ms square pulse, (e) a negative 1 ms Gaussian pulse, (f) a positive 1 ms triangular pulse.  Here, pulses deemed ``positive'' start at a maximally squeezed level and show an increase in noise up to shot noise, while for the ``negative'' pulses the measured noise starts at the shot noise and then drops to the maximum squeezing level. The desired profile can be reliably reproduced in the measured spectral noise power by calibrating the effect of the magnetic field on the squeezed noise levels. We can use this to determine the transfer function and thus calculate the magnetic field pulses required to produce the desired noise pulse shapes.

We  see   an  excellent
mapping  from the desired  to observed  pulse shapes,  importantly with no
degradation of  the maximally squeezed noise quadratures, with the
lowest noise levels recorded (maximum squeezing) always occurring at
zero magnetic field.  The applied magnetic field is chosen so that the
quadrature noise moves between the maximally squeezed level ($\approx -2.3$ dB)
and the shot noise
level, following the pulse shape we desire.  The pulses are smooth and
continuous and we have easy control over their duration and repetition
rate.  
%This  method can be  expanded to any other arbitrary pulse
%shapes (within the  detection VBW and RBW) by using the known relationship
%between squeezing and magnetic field.  We note that the squeezed noise
%levels appear to react to changes in the magnetic field on atomic
%timescales, and that this setup is primarily limited only by the bandwidth
%of our current source.

\begin{figure}[h]
	\begin{center}
	\includegraphics[width=0.4\textwidth]{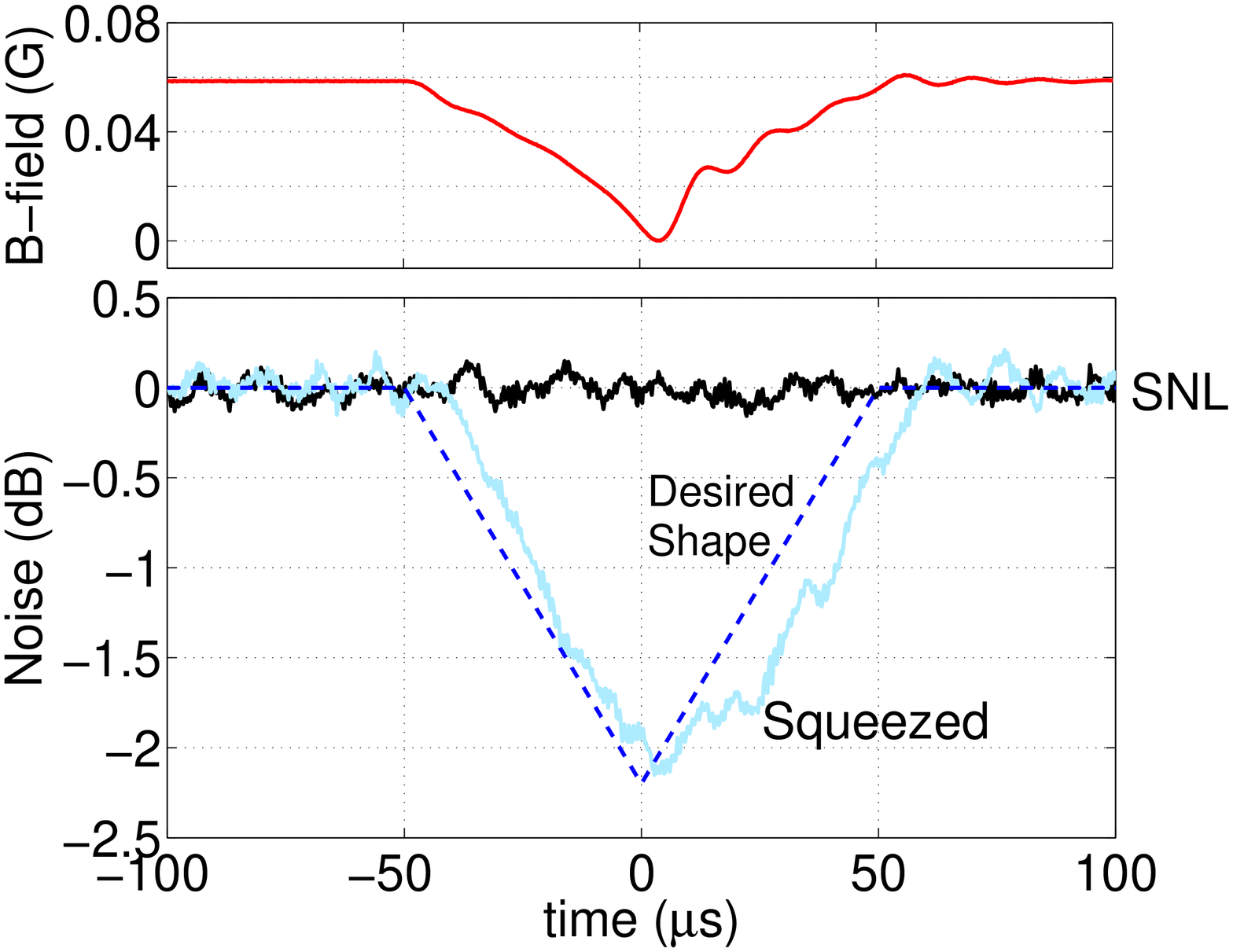}
	\end{center}
	\caption{A $100 \mu s$ triangular pulse of squeezed noise (same coding
	conventions as in Figure~\ref{fig:pulses}).  Limits of the
	current supply bandwidth cause visible oscillations in the magnetic field which
	shows up in squeezed spectrum.}
	\label{fig:ripple}
\end{figure}

The shortest  generated and detected pulses are on the  order of
30~$\mu s$, which is limited by some electronics available for the experiment. During the measurements of the squeezing  pulses  above, we set  the  spectrum analyzer  central
frequency  to  1~MHz,  the  resolution  bandwidth (RBW)  to  100~kHz,  and  video
bandwidth (VBW)  to 3~MHz while  we monitored the
time-dependent noise level with an oscilloscope. The SA bandwidth setting naturally
limits the  maximum bandwidth of  the pulses or shortest  possible detected
pulse. However,  the main  limitation on pulse duration was set  by the  homemade controllable
current source  which was used to control the solenoid current.  The bandwidth  of this
current source was limited to about 10~kHz and for shorter pulses, it distorted the programmed pulse shapes by adding unwanted transient effects.  We can see the ripples of the
set  current  in  Fig.~\ref{fig:ripple}  when
the signal  bandwidth exceeds the instrumental  one. Note however that the detected squeezing accurately follows the distorted magnetic pulse shape, illustrating that it can potentially be modulated much faster.  To  avoid  ring-down  oscillations,
 we  smooth the sharp  fronts of  the input rectangular  pulse (see
Fig.~\ref{fig:pulses}(d)).

\section{Conclusion}

In summary, we have studied the  effect of an external magnetic field on the PSR  vacuum squeezing  generated in a hot ${}^{87}$Rb vapor. We found that the longitudinal magnetic field degrades the squeezing, increasing detected noise level in the squeezed quadrature roughly linearly with applied magnetic field strength (for small fields), but without changing this quadrature angle with respect to the local oscillator. We demonstrated that we can use this dependence to generate pulses of  the squeezed vacuum with arbitrary temporal profiles. The advantages of the proposed pulsed squeezing generation method is its simplicity and robustness, good matching of the squeezing parameters to narrow coherent atomic resonances (like EIT or Raman transitions),  and the possibility to use the pump field as a local oscillator for perfect spatial mode matching. This technique is thus suitable for time encoding  of the  quantum  states for  quantum communication  and memory applications.

\section{Acknowledgments}

The authors thank Gleb Romanov and Robinjeet Singh for their
assistance with the experiment and for useful discussions. This
research was supported by NSF grant PHY-0758010.

\section*{References}
\bibliographystyle{unsrt}
\bibliography{bibliography}

\end{document}